\documentclass[crop, oneside, sn-nature]{sn-jnl}

\usepackage{graphicx,amsmath,amssymb,amsfonts,amsthm,physics,mathrsfs}

\usepackage{bbm}

\usepackage{mlmodern}

\usepackage{algpseudocode}

\theoremstyle{thmstyleone}%
\newtheorem{theorem}{Theorem}
\newtheorem{lemma}{Lemma}

\theoremstyle{definition}

\newcommand{\sset}[1]{\mathscr{#1}}
\newcommand{\graph}[1]{\mathscr{#1}}
\DeclareMathOperator{\sgn}{sgn}
\newcommand{\mdII}{V${}_2$}
\newcommand{\IIind}{{\text{V}_2}}

\begin{document}

\title{Relaxation-based dynamical Ising machines for discrete tomography}

\author{\fnm{Mikhail} \sur{Erementchouk}}\email{merement@gmail.com}
\author{\fnm{Aditya} \sur{Shukla}}\email{aditshuk@umich.edu}
\author{\fnm{Pinaki} \sur{Mazumder}}\email{pinakimazum@gmail.com}
\affil{\orgdiv{Department of Electrical Engineering and Computer Science},
  \orgname{University of Michigan}, \orgaddress{\city{Ann Arbor},
    \postcode{48104}, \state{MI}, \country{USA}}}


\abstract {
  Dynamical Ising machines are continuous dynamical systems that evolve
  from a generic initial state to a state strongly related to the ground
  state of the classical Ising model. We show that such a machine driven by
  the V${}_2$ dynamical model can solve exactly discrete tomography
  problems about reconstructing a binary image from the pixel sums along a
  discrete set of rays. In contrast to usual applications of Ising
  machines, targeting approximate solutions to optimization problems, the
  randomly initialized V${}_2$ model converges with high probability
  ($P_{\mathrm{succ}} \approx 1$) to an image precisely satisfying the
  tomographic data. For the problems with at most two rays intersecting at
  each pixel, the V${}_2$ model converges in internal machine time that
  depends only weakly on the image size. Our consideration is an example of
  how specific dynamical systems can produce exact solutions to highly
  non-trivial data processing tasks. Crucially, this solving capability
  arises from the dynamical features of the V${}_2$ model itself, in
  particular its equations of motion that enable non-local transitions of
  the discrete component of the relaxed spin beyond Hamming-neighborhood
  constraints, rather than from merely recasting the tomography problem in
  spin form. }

\keywords{dynamical computations, combinatorial optimization, dynamical Ising
  machines, spin systems, discrete tomography, non-local transitions}

\maketitle

\section{Introduction}

Ising machines represent an approach to computing that utilizes inherent
properties of networks of classical spins, taking values $\sigma = \pm1$.
These computational properties attract researchers' attention for a long
time~\cite{kirkpatrickOptimization1983, hopfieldNeurons1984,
  cernyThermodynamical1985, fuApplication1986}, and stem from two key
properties of the Ising model ground state, the spin
distribution with the lowest energy. First, the ground state can be
regarded as solving a quadratic unconstrained binary optimization
problem~\cite{kochenbergerBinary2013}. Second, by associating the spin
distribution in the ground state with a partition of the graph nodes, one
obtains the maximum (weighted) cut of the model
graph~\cite{barahonaComputational1982}. Finding the maximum cut is an
NP-complete problem~\cite{karpReducibility1972,gareySimplified1976}, which
puts the Ising model into the general computing
perspective~\cite{lucasIsing2014}.

The equivalence between the ground state and the max-cut problems follows
from the following observation. Let the network of $N$ classical spins be
defined on a graph described by a weighted adjacency matrix with the matrix
elements $A_{m,n}$. Regarding spin configuration (the distribution of spins
on the graph)
$\boldsymbol{\sigma} = \left( \sigma_1, \ldots, \sigma_N \right)$ as a system of interacting
spins as in the classical Ising model, we associate with
$\boldsymbol{\sigma}$ an Ising energy
\begin{equation}\label{eq:Ising-H}
  \mathcal{H}(\boldsymbol{\sigma}) = \frac{1}{2} \sum_{m,n} A_{m,n} \sigma_m \sigma_n.
\end{equation}

Alternatively, the spin configuration can be regarded as a designation of
partition for individual graph nodes. For example, nodes with
$\sigma_n = 1$ and $\sigma_n = -1$ are considered belonging to partitions
$1$ and $2$, respectively. This defines a partition of the graph nodes that
can be characterized by the weight of all cut edges, that is edges that
connect nodes in different partitions. Introducing the cut indicator
function $\chi(m,n) = (1 - \sigma_m \sigma_n)/2$ for edge $(m,n)$, we obtain the cut
function describing the spin configuration
\begin{equation}\label{eq:cut-function}
  \mathcal{C}(\boldsymbol{\sigma}) = \frac{1}{4} \sum_{m,n} A_{m,n} \left( 1 - \sigma_m \sigma_n \right).
\end{equation}
Functions $\mathcal{H}(\boldsymbol{\sigma})$ and $\mathcal{C}(\boldsymbol{\sigma})$ are obviously related
$\mathcal{C}(\boldsymbol{\sigma}) = \sum_{m,n} A_{m,n} / 4 - \mathcal{H}(\boldsymbol{\sigma}) / 4$. Thus,
indeed, finding the maximum cut partition, configuration $\boldsymbol{\sigma}$
yielding the maximum of $\mathcal{C}(\boldsymbol{\sigma})$, is
equivalent to finding the ground state, configuration $\boldsymbol{\sigma}$, at
which $\mathcal{H}(\boldsymbol{\sigma})$ takes the minimum value.

In dynamical Ising machines~\cite{mohseniIsingMachinesHardware2022}, the
computational properties of classical spin networks are exposed through
implementing the networks in various continuous dynamical systems designed
or selected with the objective that the system progresses towards a state
that is tightly related to the network ground state or, equivalently, to
the maximum cut partitioning of the network graph. This is achieved by
representing the binary spins by continuous variables,
$\sigma_m \to \xi_m$, and associating with the collective state of the dynamical
variables a Lyapunov, or more general objective function. The choice of
function is constrained by two requirements. First, it should change
monotonously on the trajectories of the dynamical system. Second, spin-like
states, with $\boldsymbol{\xi} = \boldsymbol{\sigma}$, yield the same value of the
objective function as its discrete counterpart. For example, if the
dynamical system is characterized by a continuous energy function,
$\mathcal{H}_M(\boldsymbol{\xi})$, then
$\mathcal{H}_M(\boldsymbol{\xi} = \boldsymbol{\sigma}) = \mathcal{H}(\boldsymbol{\sigma})$.

Requiring that the objective function monotonously changes with evolution
puts Ising machines into the context of solving optimization problems,
which, at present, is a dominating approach to Ising machines. However, the
optimization perspective is too broad, and the discrete optimization is
famous for ubiquitous hard problems~\cite{papadimitriouCombinatorial1998,
  aroraComputational2009}. For example, the general max-cut problem, which
is native to Ising machines, besides being NP-hard is also
APX-hard~\cite{gareyComputersIntractabilityGuide1990}, meaning that there
is a limit to approximation that can be achieved in polynomial time.
Consequently, the Ising machines are commonly regarded as heuristic
optimization engines aiming for approximate solutions. This essentially
puts Ising machines out of the context of solving ``common'' information
processing tasks that demand exact solutions, even if non-deterministic.

In the present paper, we demonstrate that individual dynamical models
driving Ising machines possess a powerful computational resource that
allows them to solve certain non-trivial data-processing problems exactly.
Specifically, we show that an Ising machine driven by the \mdII{} dynamical
model~\cite{erementchoukSelfcontained2023} can solve discrete tomography
problem about restoring a binary image from its tomographic data, the
collection of pixel sums along a discrete set of rays. Notably, this
ability stems from specific dynamical features of the \mdII{} model, in
particular, the ability to traverse spin configurations transcending
Hamming-bound constraints.

Our consideration shows that Ising machines can be considered in a much
broader context as platforms for realizations  of an alternative class of
viable algorithms performing ``traditional'' data-processing tasks.

The rest of the paper is organized as follows. In
Section~\ref{sec:v2--model}, we overview the main properties of the \mdII{}
model. In Section~\ref{sec:dt-graph-formulation}, we provide the
graph-theoretical formulation of the discrete tomography problem to solve
by the \mdII{} model. In Section~\ref{sec:convergence}, we analyze the
convergence of the \mdII{} model to the solutions of the tomography problem.



% ...


\section{Basic properties of the V${}_2$ model}
\label{sec:v2--model}

The \mdII{} model is a representative of a broad class of dissipative
dynamical systems driving relaxation-based Ising machines.

The \mdII{} model describes a special class of dynamical systems with
dissipative dynamics. The dynamical variables of the system are associated
with nodes $\sset{V}$ of undirected finite graph
$\graph{G} = \left\{ \sset{V}, \sset{E} \right\}$, whose set of edges
$\sset{E}$ with assigned weights (both, positive and negative weights are
allowed) represent coupling between the variables. We will denote the
distribution of variables associated with the graph nodes by
$\boldsymbol{\xi} = \left( \xi_1, \ldots, \xi_N \right)$. The evolution of the system
is governed by equations of motion that are convenient to define as 
$\dot{\boldsymbol{\xi}} = 2 \nabla_{\boldsymbol{\xi}} \mathcal{C}_{\IIind}(\boldsymbol{\xi})$,
ensuring that $\mathcal{C}_{\IIind}(\boldsymbol{\xi})$ is a non-decreasing function of
time. The \mdII{} model is defined by
\begin{equation}\label{eq:mdII-objective}
  \mathcal{C}_{\IIind}(\boldsymbol{\xi}) = \frac{1}{4}
  \sum_{\alpha, \beta \in \sset{V}} A_{\alpha, \beta} \abs{\xi_\alpha - \xi_\beta}_{[-2,2]},
\end{equation}
where $A_{\alpha, \beta}$ is the (weighted) adjacency matrix of graph
$\graph{G}$, and $\abs{\ldots}_{[-2,2]}$ is a periodic function with period
$4$ defined for $-2 \leq \xi \leq 2$ by $\abs{\xi}_{[-2,2]} = \abs{\xi}$.

Taking $\xi_\alpha = \pm 1$ turns $\mathcal{C}_{\IIind}(\boldsymbol{\xi})$ into the total weight
of edges between the nodes corresponding to $\xi$'s with different signs. In
other words, $\mathcal{C}_{\IIind}(\boldsymbol{\xi})$ is a relaxation of the cut
function. In turn, since the system is either at rest, or
$\mathcal{C}_{\IIind}(\boldsymbol{\xi})$ increases with time, the the \mdII{} model
evolution can be related to finding the maximum cut of graph $\sset{G}$.

\begin{figure}[tb]
  \centering
  \includegraphics[width=3.0in]{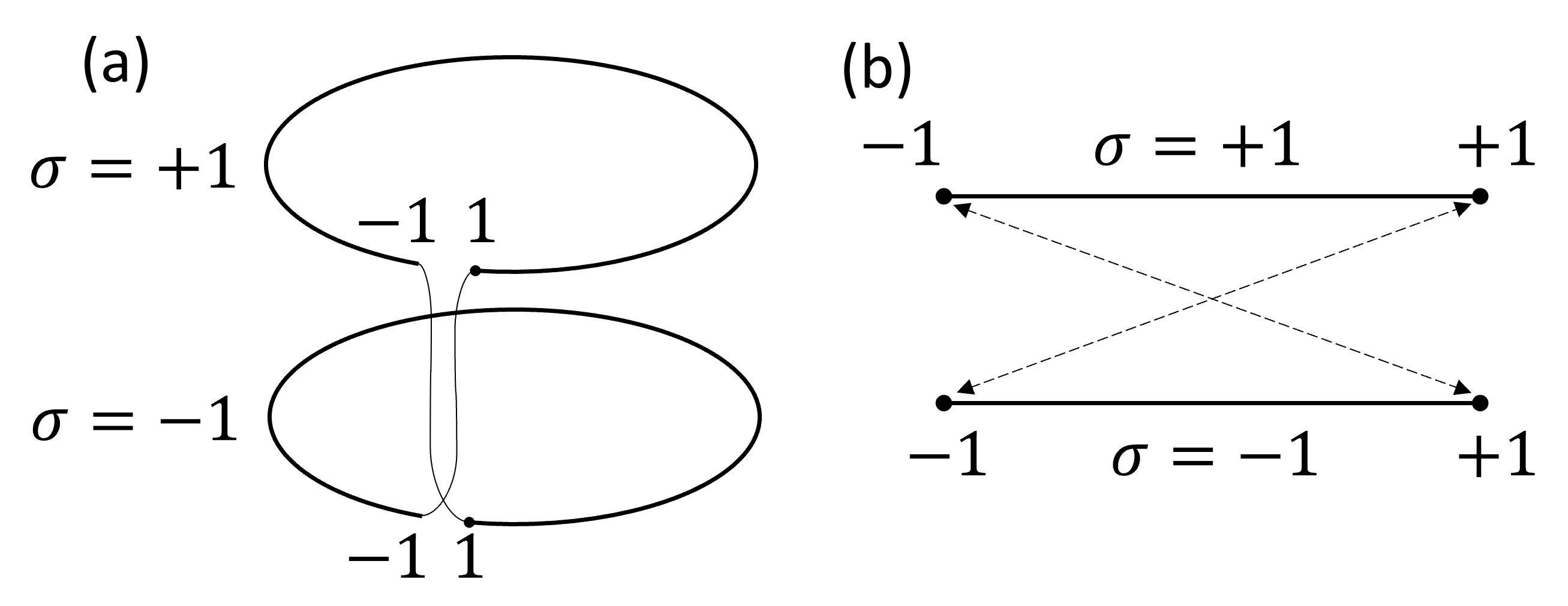}
  \caption{The phase space of relaxed spins in the \mdII{} model as (a) two
    circles with the transition rule, or, alternatively, as (b) two
    intervals $[-1, 1]$ with respectively identified boundary points.}
  \label{fig:spin_phase_space}
\end{figure}

The connection with the combinatorial structures emerging on the graph and
the \mdII{} model is made more transparent by employing the periodicity of
the kernel in Eq.~\eqref{eq:mdII-objective} and introducing new dynamical
variables, relaxed spins, by the relation
$\xi_\alpha = \sigma_\alpha + X_\alpha + 4k_a$. Here,
$\sigma \in \left\{ -1, 1 \right\}$ and $X \in (-1, 1]$ are the binary and
continuous components of the relaxed spin, respectively. Integer $k_\alpha$
accounts for the kernel periodicity in
$\mathcal{C}_{\IIind}(\boldsymbol{\xi})$ and plays no role in the following. Rewriting
Eq.~\eqref{eq:mdII-objective} using the relaxed spin representation, we
obtain
\begin{equation}\label{eq:cut-v2}
  \mathcal{C}_{\IIind}(\boldsymbol{\sigma}, \mathbf{X}) = \mathcal{C}(\boldsymbol{\sigma}) +
  \widetilde{\mathcal{C}}_{\IIind}(\boldsymbol{\sigma}, \mathbf{X}),
\end{equation}
with
\begin{equation}\label{eq:cut-v2-add}
  \widetilde{\mathcal{C}}_{\IIind}(\boldsymbol{\sigma}, \mathbf{X}) =
  \frac{1}{4} \sum_{\alpha, \beta} A_{\alpha, \beta} \sigma_\alpha \sigma_\beta \abs{X_\alpha - X_\beta}.
\end{equation}
Respectively, the dynamics of relaxed spins is governed
for $X_\alpha \in [-1, 1)$ by
\begin{equation}\label{eq:eqm-v2-x}
  \dot{X}_\alpha = \sum_{\beta \in \sset{V}} A_{\alpha,\beta} \sigma_\alpha \sigma_\beta \sgn\left( X_\alpha - X_\beta \right),
\end{equation}
where $\sgn(X)$ is the sign function with the adopted convention
$\sgn(0) = 0$. The evolution described by these equations occurs on the
phase space of relaxed spins illustrated by
Fig.~\ref{fig:spin_phase_space}. The topology of the phase space is
reflected by supplementing the equations of motion with the rule of
negotiating the basepoint in the wedge sum by the phase point: whenever
$X_\alpha$ crosses the $X = \pm1$ boundary, $\sigma_\alpha$ changes sign.

The computational significance of the evolution of dissipative dynamical
Ising machines can be established by relating it to the properties of the
terminal states of the machine initiated in a generic state. The dynamics
governed by Eqs.~\eqref{eq:eqm-v2-x} terminates at critical points of
$\mathcal{C}_{\IIind}(\boldsymbol{\sigma}, \mathbf{X})$ that possess several general key
features discussed in~\cite{erementchoukSelfcontained2023}).

First of all, while the terminal states of the \mdII{} model are not
necessarily binary (the continuous components of the relaxed spins may not
vanish in the terminal state), it is trivial to round them optimally
because of the vanishing term
$\widetilde{\mathcal{C}}_{\IIind}(\boldsymbol{\sigma}, \mathbf{X})$ at critical points.
The relaxed cut function
$\widetilde{\mathcal{C}}_{\IIind}(\boldsymbol{\sigma}, \mathbf{X})$ is invariant with
respect to global rotations of the phase circles, or, alternatively, global
homogeneous translation $\mathbf{X} \to \mathbf{X} + r \mathbf{1}$, with
$\mathbf{1} = \left( 1, 1, \ldots, 1 \right)$, performed on the model phase
space with the respective inversions of the binary component of the relaxed
spin. Consequently, the binary cut function
$\mathcal{C}(\boldsymbol{\sigma})$ evaluated in terminal states of \mdII{} model does not
depend on the choice of the (global) origin on the model phase circles.
While rotating the phase circles may produce different binary states
$\boldsymbol{\sigma}$, they all correspond to the same value of the cut function
and, from this perspective, are equivalent. Hence, nonbinary solutions can
be mapped to binary states by simply ignoring the continuous component of
the relaxed spin. It is worth noting that this is a specific property of
the \mdII{} model. Various relaxations can be productively represented in
terms of relaxed spins with the relaxed cut function invariant with respect
to global phase rotations. However, the binary states produced by such
rotations may yield different cut values raising the special problem of
optimal rounding~(see, e.g., Section~8.2 in~\cite{punnenQuadratic2022}).

The terminal states of the \mdII{} model have a characteristic structure
with relaxed spins grouping into clusters with the same value of the
continuous component. This is a consequence of the fact that for states
with all continuous component different, the relaxed cut function is a
linear function of $\mathbf{X}$ with the vanishing Hessian.
To account for this
circumstance, we introduce the notion of \emph{strong clusters}. Critical
point $(\boldsymbol{\sigma}, \mathbf{X})$ partitions the set of graph nodes
$\sset{V}$ into a system of $1 \leq S \leq \abs{\sset{V}}$ disjoint subsets
(strong clusters)
$\sset{V} = \bigcup_{p = 1}^S \sset{V}^{(p)}(\boldsymbol{\sigma}, \mathbf{X})$ with
$\sset{V}^{(p)}(\boldsymbol{\sigma}, \mathbf{X}) = \left\{ \alpha \in \sset{V} : X_\alpha =
  X^{(p)} \right\}$ and $X^{(p)} \ne X^{(p')}$ for $p \ne p'$. It is convenient
to enumerate clusters in such a way that $X^{(p)} > X^{(p')}$ for $p > p'$.

The convergence to a system of strong clusters does not constrain the
structure of the terminal states since the definition does not preclude
clusters from having only one node. Thus, an arbitrary distribution
$\mathbf{X}$ can be regarded as a system of strong clusters. However,
commonly, for instance while solving the discrete tomography problem, in
terminal states, all relaxed spins group into a few strong clusters.

Coinciding coordinates in strong clusters require addressing the problems
related to discontinuities in the right-hand-side of
Eq.~\eqref{eq:eqm-v2-x}. We deal with these difficulties using the
separability of strong clusters emerging from generic initial conditions
and continuity and boundedness of the relaxed cut function. The character
of interaction between the relaxed spins in the \mdII{} model allows one to
separate the dynamical contributions stemming from the interactions within
the clusters and between the relaxed spins in different clusters, which
enables an analysis of the convergence to a system of strong clusters and
its stability. To utilize this feature, we introduce \emph{weak clusters},
when groups of relaxed spins can be clearly identified without having all
continuous components within a group coinciding. More formally, a system of
{weak clusters} is a partition
$\sset{V} = \bigcup_p \sset{U}^{(p)}(\boldsymbol{\sigma}, \mathbf{X})$ induced by a
system of non-intersecting open intervals
$\sset{I}^{(p)} = \left( X^{(p)}_-, X^{(p)}_+ \right) \subset [-1, 1)$, such that
$\sset{U}^{(p)}(\boldsymbol{\sigma}, \mathbf{X}) = \left\{ \alpha \in \sset{V} : X_\alpha \in
  \sset{I}^{(p)} \right\}$.
We say that a weak cluster is \emph{generic} if all continuous components
of the relaxed spins are distinct: $X_\alpha \ne X_\beta$ for all
$\alpha, \beta \in \sset{V}$ such that $\alpha \ne \beta$.

The next distinctive property of the relaxed cut function
$\mathcal{C}_{\IIind}(\boldsymbol{\sigma}, \mathbf{X})$ is that it has no local maxima:
its critical points are either saddle points or max-cut. At the same time,
the evolution of the \mdII{} model can terminate in a saddle point.
Identifying such situations is the main problem in investigating the
convergence of the \mdII{} model.

In this regard, another important feature of the \mdII{} model is worth
emphasizing as it greatly enhances the model computational power. While
departing from a slight perturbation of a binary state, the system of
relaxed spins ends up in a state with the cut not smaller than that of the
initial state. More precisely, if the initial state is
$\left( \boldsymbol{\sigma}(0), \mathbf{X}(0) \right)$, the \mdII{} model
progresses in such a way that
$\mathcal{C}(\boldsymbol{\sigma}(t)) \geq \mathcal{C}(\boldsymbol{\sigma}(0))$. Moreover, if
$\boldsymbol{\sigma}(0)$ does not yield the maximum cut,
$\mathcal{C}(\boldsymbol{\sigma}(0)) < \overline{\mathcal{C}_{\graph{G}}}$, the probability of
improving cut by a random perturbation of the continuous component is
strictly positive (Theorem~6 in~\cite{erementchoukSelfcontained2023}). We
will call the randomly agitated \mdII-machine a system, whose dynamics is
described by the \mdII{} model and whose terminal states are randomly
agitated.

The beneficial property of random agitations is their ability to break
disadvantageous patterns that may emerge due to a peculiar character of the
initial condition. For example, the model initialized in state
$\left( \boldsymbol{\sigma}, \mathbf{0} \right)$ with random
$\boldsymbol{\sigma}$ will stay in this computationally insignificant state. The
same trivial result is obtained for an ``agitated'' initial state of the
form $\left( \boldsymbol{\sigma}, \kappa \mathbf{1} \right)$, with
$\abs{\kappa} < 1$. Random agitations eliminate such pathological effects.


\section{Graph formulation of discrete tomography}
\label{sec:dt-graph-formulation}

The main distinguishing features of the dynamics of the \mdII{} model are
established in relation to the evolution of a system of
relaxed spin on graphs toward their ground state. Therefore,
the discussion of how the \mdII{} model solves the discrete tomography
problem significantly benefits from reformulating the latter in
graph-theoretical terms.

The discrete tomography envelops a great variety of problems related to
reconstructing images from limited information~\cite{hermanDiscrete1999,
  hermanAdvancesDiscreteTomography2007}. Here, we focus on the canonical
problem of recovering a binary image from its projections along a selected
family of rays. Usually, this problem is formulated as follows. The binary
image is represented by $\hat{s}$, a square $W \times W$ matrix with elements
$s_{m,n} \in \left\{ 0, 1 \right\}$. The tomographic data is given by two
vectors $\mathbf{P}^{(\mathrm{row})}$ (assigning to each row the sum over
columns) and $\mathbf{P}^{(\mathrm{col})}$ (assigning to each column the
sum over rows) with components $P^{(\mathrm{row})}_m = \sum_n s_{m,n}$ and
$P^{(\mathrm{col})}_n = \sum_m s_{m,n}$. Solving a tomography problem means
reconstructing the binary image from the tomographic data: finding such
distribution of set pixels in the $W \times W$ matrix, whose projections along
columns and columns reproduces the tomographic data. It should be noted
that, in general, the tomography problem is ill-posed: vectors
$\mathbf{P}^{(\mathrm{row})}$ and $\mathbf{P}^{(\mathrm{col})}$ do not
necessarily define $\hat{s}$ uniquely~\cite{ryserCombinatorial1987,
  galeTheorem1987, hermanDiscrete1999}. For example, when
$\mathbf{P}^{(\mathrm{row})}$ have coinciding elements, say,
${P}^{(\mathrm{row})}_m = {P}^{(\mathrm{row})}_{m'}$ for some $m \ne m'$,
then the image obtained by permuting rows $m$ and $m'$ also satisfies the
same tomographic data but may be different. Since the objective of the
present paper is to outline the computational capabilities of the
\mdII-machine, we will understand the tomography problem as finding
\emph{any} image $\hat{s}$ that satisfies the tomographic data. From this
perspective, it is more important that the tomographic data is
\emph{consistent}: there exists at least one image satisfying the data. In
what follows, we will assume that this is indeed the case.

Another important condition that we will assume always fulfilled is that
the tomographic data is collected along such a system of rays that any two
rays may intersect at most at one point. For the example above, this
condition is met trivially because each pixel $s_{m,n}$ contributes only to
$P^{(\mathrm{row})}_m$ and $P^{(\mathrm{col})}_n$. Consequently, two
distinct pixels $s_{m,n}$ and $s_{m',n'}$ contribute to the same
tomographic data only if $m = m'$, $n \ne n'$ or $m \ne m'$, $n = n'$.

To reformulate the tomography problem in graph-theoretical terms, we
identify the image pixels with with set $\sset{V}$ of nodes and
define the binary image as a function on the nodes:
$\mathbf{s} : \sset{V} \to \left\{ 0, 1 \right\}$. Next, the system of rays,
$\sset{R} = \left\{ \sset{R}(r) \subset \sset{V}, 1 \leq r \leq R : \abs{\sset{V}(r) \cap
    \sset{V}(d')} \leq 1, r \ne r' \right\}$, is a set of rays
$\sset{R}(r)$, subsets of nodes such that two different subsets have at
most one node in common.

The \emph{tomographic data} is a function
$\mathbf{P}_{\mathbf{s}} : \sset{R} \to \mathbb{Z}$ evaluating the pixel-sum along the
ray, the total weight of intersection of image $\mathbf{s}$ with the rays,
$\mathbf{P}_{\mathbf{s}} : \sset{R}(r) \mapsto P_{\mathbf{s}}(r) := \sum_{u \in
  \sset{R}(r)} s(u)$. In other words, $P_{\mathbf{s}}(r)$ is the number of
set pixels in the $r$-th ray traversing image ${\mathbf{s}}$. Respectively,
the \emph{tomography problem} is posed as finding a binary image
$\mathbf{s}' : \sset{V} \to \left\{ 0, 1 \right\}$ with matching tomographic
data, $\mathbf{P}_{\mathbf{s}'} = \mathbf{P}_{\mathbf{s}}$.

In these terms, the canonical formulation corresponds to a set of $W^2$
nodes arranged as a $W \times W$ square on a 2D integer lattice, with subsets
$\sset{R}(r)$ representing the rows and the columns of the square. At the
same time, the generalized formulation covers other classes of besides
reconstructing 2D images given the tomographic data collected along the
rows and columns of the image. For example, clues specifying pixels that
must be set or unset in the reconstructed image can be defined as a
tomographic data collected over single-pixel rays. Furthermore, the
generalized reformulation naturally extends to problems
with a more complex geometry, such as reconstructing 3D images (see
Section~\ref{sec:3d-pcb} below).

Since the ground state of a spin network is equivalent to a maximum cut
partition, we seek for a graph, whose maximum cut corresponds
reconstructing the binary image satisfying the tomographic data. To this
end, we consider a particular ray $\sset{R}(r) \in \sset{R}$ with
$N(r) = \abs{\sset{R}(r)}$ nodes and its partitioning given by the spin
configuration $\boldsymbol{\sigma}$. Associating spins up with set pixels, we
find that the reconstructed single-ray image satisfies the tomographic data
if
$Q(r; \boldsymbol{\sigma}) := \sum_{\alpha \in \sset{R}(r)} \sigma_{\alpha} + \widetilde{q}_{0}(r)
\sigma_{0} = 0$, where $\widetilde{q}_{0}(r) = N(r) - 2 P_{\mathbf{s}}(r)$ is
the tomographic data written for the spin representation of the binary
image and $\sigma_{0} \equiv 1$ is the fixed auxiliary spin. The deviation of the
image represented by $\boldsymbol{\sigma}$ from the tomographic data can be
quantified by penalty $Q^{2}(r; \boldsymbol{\sigma})$, which has a direct
combinatorial meaning. With ray $\sset{R}(r)$, we associate graph
$\widetilde{\graph{K}}(r)$, a complete graph $\graph{K}_{\sset{R}(r)}$ on
$\sset{R}(r)$ with added dominating vertex (auxiliary spin)
$\left\{ 0 \right\}$ connected to the rest of the nodes by edges with
weight $\widetilde{q}_{0}(r)$. We denote by
$\widetilde{\sset{R}}(r) = \sset{R}(r) \cup \left\{ 0 \right\}$ the set of
nodes in ray $r$ with added dominating vertex. Configuration
$\boldsymbol{\sigma}$ induces cut of $\widetilde{\graph{K}}(r)$ that can be
written as
$ \mathcal{C}(\boldsymbol{\sigma})_{\widetilde{\graph{K}}(r)} =
\overline{\mathcal{C}}_{\widetilde{\graph{K}}(r)} -
\frac{1}{4} Q^{2}(r; \boldsymbol{\sigma})$,
where
$\overline{\mathcal{C}}_{\widetilde{\graph{K}}(r)} = \left( N(r) -
  P_{\mathbf{s}}(r)\right)^{2}$ is the maximum cut of
$\widetilde{\graph{K}}(r)$.

\begin{figure}[tb]
  \centering
  \includegraphics[width=3in]{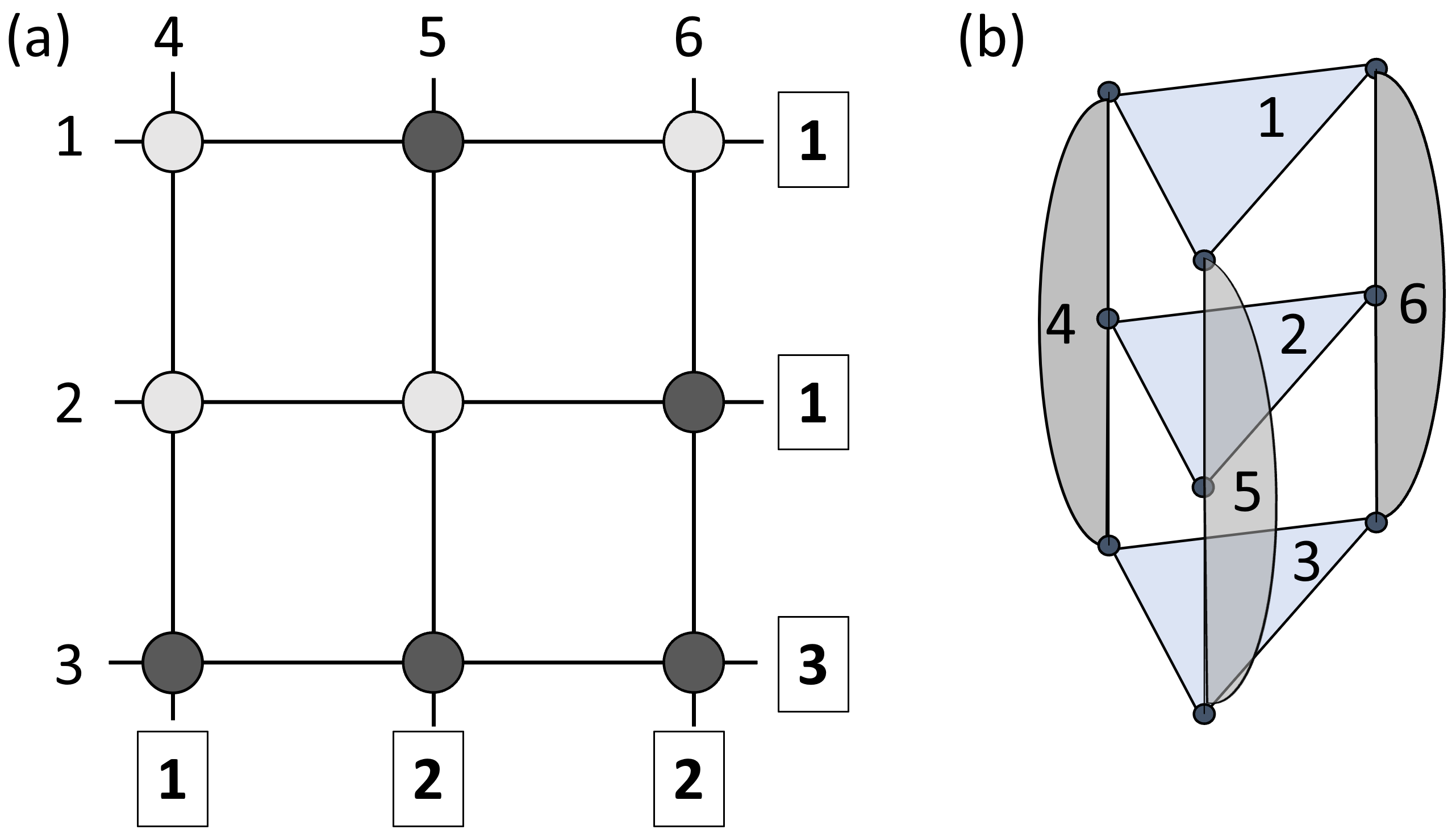}
  \caption{An example of a graph representation of a discrete tomography
    problem. (a) A tomography problem set on $3\times3$ grid. Circles represent
    pixels, horizontal and vertical lines indicate rays, along which the
    tomographic data is collected, plain numbers are an enumeration of
    rays, bold framed numbers are tomographic data for the example image (a
    glider from Conway's Game of Life~\cite{gardnerWheels1983}) with dark
    shaded circles representing pixels set ON. (b) A graph corresponding to
    problem shown in~(a). Enumerated shaded regions show cliques
    representing the corresponding rays and contain auxiliary nodes holding
    the auxiliary spin.}
  \label{fig:graph-form}
\end{figure}

Extending such consideration to all rays, we find that reconstructing
binary image from tomographic data is equivalent to finding the maximum cut
partition of graph
$\graph{G}_{\mathbf{s}, \mathbf{P}} = \left\{ \widetilde{\sset{V}},
  \sset{E} \right\}$, with
$\widetilde{\sset{V}} = \sset{V} \cup \left\{ 0 \right\}$, constructed by the
union of cliques $\widetilde{\graph{K}}(r)$ with $r = 1, \ldots, R$. An example
of the graph representing a discrete tomography problem on a $3 \times 3$ grid
is shown in Fig.~\ref{fig:graph-form}.

Depending on the context, it may be convenient either to regard a set of
individual auxiliary spins associated with different rays, or to contract
all the auxiliary spins to a single dominating vertex. During such a
contraction, multiple (or parallel) edges may form because a single node
may belong to several rays and hence can be connected to multiple auxiliary
nodes corresponding to different cliques. Such multiple edges, in turn, are
contracted to a single edge with the accumulated weight. For example, let
node $\alpha$ belong to rays $\sset{R}\left( r_1 \right)$ and
$\sset{R}\left( r_2 \right)$ with the respective
$\widetilde{q}_{0}\left( r_{1} \right) $ and
$\widetilde{q}_{0}\left( r_{2} \right) $, then in
$\graph{G}_{\mathbf{s}, \mathbf{P}}$ with contracted auxiliary spins, node
$\alpha$ is connected to the dominating vertex by the edge with weight
$\widetilde{q}_{0}\left( r_{1} \right) + \widetilde{q}_{0}\left( r_{2}
\right)$.

On the other hand, to find the combinatorial meaning of a solution to the
discrete tomography problem, it is convenient to keep the auxiliary spins
separated because in this case cliques $\widetilde{\graph{K}}(r)$ do not
share edges. Consequently, the cut weight produced by partitioning
$\graph{G}_{\mathbf{s}, \mathbf{P}}$ is the sum of cut weights produced by
induced partitioning of $\widetilde{\graph{K}}_{\sset{R}(r)}$, yielding
\begin{equation}\label{eq:cut_tom_graph}
  \mathcal{C}_{\graph{G}_{\mathbf{s}, \mathbf{P}}}(\boldsymbol{\sigma}) = \sum_r
  \mathcal{C}_{\widetilde{\graph{K}}(r)}(\boldsymbol{\sigma}) =
  \overline{\mathcal{C}}_{\graph{G}_{\mathbf{s}, \mathbf{P}}} -
  \frac{1}{4} \sum_r Q^{2}(r; \boldsymbol{\sigma}),
\end{equation}
where
$\overline{\mathcal{C}}_{\graph{G}_{\mathbf{s}, \mathbf{P}}} = \sum_{r} \left(N(r) -
  P_{\mathbf{s}}(r) \right)^{2}$ is the maximum cut of
$\graph{G}_{\mathbf{s}, \mathbf{P}}$. Here, we observe that the
partitioning yielding the maximum cut of
$\graph{G}_{\mathbf{s}, \mathbf{P}}$ simultaneously delivers the maximum
cut of all cliques $\widetilde{\graph{K}}_{\sset{R}(r)}$. This is a
graph-theoretical formulation of the consistency of the tomographic data,
which plays the key role in the following.


Once a problem is formulated as a max-cut problem for a graph with weighted
edges, the equations of motion of the \mdII{} model can be obtained from
graph weighted adjacency matrix using Eq.~\eqref{eq:eqm-v2-x}. Such an
approach may be useful for, say, generating equations of motion for
numerical simulations. However, for a productive analysis of the
computational capabilities of the \mdII{} model, it is constructive to
reflect the structure of the graph representing the tomography problem
directly in the equations of motion. Therefore, we delay writing the
explicit form of these equations till the next section [see
Eqs.~\eqref{eq:eqm-v2-q-multi} and~\eqref{eq:e_field-def-multi}], where
they will be presented in a much more compact form compared to a
straightforward explication of Eq.~\eqref{eq:eqm-v2-x}.

\section{Convergence to the solutions}
\label{sec:convergence}

\subsection{Terminal states for single-ray problems}
\label{sec:terminal-states-tomo}

The computational capabilities of an Ising machine are determined by the
properties of terminal states that can be reached with high probability
starting from a generic initial state. As discussed above, the only stable
critical points of the \mdII{} model are those yielding the maximum cut of
the network, and, strictly speaking, there are no local maxima. However,
owing to the structure of the equations of motion, the dynamical system
governed by the \mdII{} model may terminate in a saddle point. The picture
significantly simplifies in view of the fact that all critical points of
$\mathcal{C}_{\IIind}(\boldsymbol{\sigma}, \mathbf{X})$ belong to critical manifolds
enumerable by purely binary states~\cite{erementchoukSelfcontained2023}.
Therefore, while considering whether a particular terminal state can be
reached, the binary component can be considered fixed. States unstable with
respect to generic small perturbations can be disregarded. In the worst
case, the system can be led away from them by random agitations. As we will
see, many saddle critical manifolds can be ruled out based on this simple
analysis. However, for some critical manifolds such consideration is not
sufficient.

The ground for a more refined analysis is the observation that stability of
a critical point is determined by the equations of motion of the same
structure as of the equations arising in 1D electrostatics with vector
charges. To relate these seemingly different situations, we consider the
simplest case when the image is covered by a single ray:
$\sset{R} = \left\{ \sset{R}(1) \right\}$ with $\sset{R}(1) = \sset{V}$,
or, in other words, graph $\graph{G}_{\mathbf{s}, \mathbf{P}}$ coincides
with clique $\widetilde{\graph{K}}_{\sset{R}(1)}$. In this case, the
tomographic data reduces to a single number $P(1)$, the number of activated
pixels in the image. With each free node $\alpha \ne 0$, we associate charge
$q_\alpha = \sigma_\alpha$, and to the auxiliary node we assign charge
$q_0 = \sigma_0\widetilde{q}(1) = N(1) - 2 P(1)$. Using these notations,
equations of motion~\eqref{eq:eqm-v2-x} for free spins can be rewritten in
a simple form
\begin{equation}\label{eq:eqm-v2-q-single}
  \dot{X}_\alpha = q_\alpha \epsilon\left( X_\alpha \right) ,
\end{equation}
where
\begin{equation}\label{eq:e_field-def-single}
  \epsilon\left( X \right) = \sum_{\beta:X_\beta < X} q_\beta - \sum_{\beta:X_\beta > X} q_\beta.
\end{equation}
Here, the summation runs over all spins, including the auxiliary one.

Next, we notice that if we are interested only in establishing the
stability of a given spin configuration $\boldsymbol{\sigma}$, the system of
equations~\eqref{eq:eqm-v2-q-single} can regarded as defined on the whole
real line. It is straightforward to show that spin distribution
$\boldsymbol{\sigma}(0)$ is stable with respect to perturbation
$\mathbf{X}(0)$, that is the system initially in state
$\left( \boldsymbol{\sigma}(0), \mathbf{X}(0) \right)$ terminates in a state
with the same discrete component, if and only if the dynamics governed by
Eqs.~\eqref{eq:eqm-v2-q-single} on $\mathbb{R}$ is bound. Consequently, we can
investigate the stability of relaxed-spin distributions characterized by
the fixed distribution of the discrete component $\boldsymbol{\sigma}$ by using
the framework of stability in the Lyapunov sense.

Finally, we observe that the stability of solutions of
Eqs.~\eqref{eq:eqm-v2-q-single} on $\mathbb{R}$ is equivalent to stability of a
system of charged particles on 1D (or a system of parallel charged planes)
with the same charges, $q_i$, as defined above for the relaxed spins, and
the particle corresponding to the auxiliary spin fixed at the origin. The
equivalence follows from the observation that, up to the choice of time
scale, the right-hand-side of the equations of motion governing the charged
particles has the same form, $q_\alpha \epsilon\left( X_\alpha \right)$, as in
Eqs.~\eqref{eq:eqm-v2-q-single} and~\eqref{eq:e_field-def-single}. The
distinguishing feature of the equations of motion for charged particles is
that they are a system of the \emph{second} order differential equations.
However, from the stability perspective, this distinction is only technical
and plays a minor role.

From this perspective, the penalty associated with missing the tomographic
constraint can be written as
  $Q(\boldsymbol{\sigma}) = \sum_{\alpha \in \widetilde{\sset{K}}(r)} q_\alpha$
and interpreted as the total ray charge. Respectively, it is constructive
to define charges of other collections of relaxed spins, such as strong and
weak clusters. For instance, we obviously have that if a weak cluster
converges to a strong cluster, then they have the same charge.

When the tomographic constraint is met, the ray charge is zero,
$Q(\boldsymbol{\sigma}) = 0$. Since the tomography problem has a solution, the
zero value is feasible. Then, it is not difficult to see that
$Q(\boldsymbol{\sigma})$ may take only even values. This, simple but important,
observation together with the analogy with stability of a system of
electric charges immediately yield a key property.

\begin{theorem} \label{thm:solution-single}
  For a single ray tomography
  problem, states with $Q(\boldsymbol{\sigma}) = 0$ (the solutions of the
  problem) are the only stable states of the \mdII{} model. Moreover,
  starting from a generic initial state, \mdII{} model terminates in a
  $Q(\boldsymbol{\sigma}) = 0$ state.
\end{theorem}

The first part of the theorem is a simple consequence of the fact that
at equilibria of the \mdII{} model, the relaxed and discrete cut
function coincide, and equilibria are either max-cut states or are saddle
points. If a terminal state is a max-cut state, then $Q{\boldsymbol{\sigma}} =
0$, and $\boldsymbol{\sigma}$ delivers a solution to the single-ray tomography
problem. It the system terminates in a saddle point, then, due to
continuity of $\mathcal{C}_{\IIind}(\boldsymbol{\sigma}, \mathbf{X})$, there exist
variations of the continuous component of the relaxed spin increasing
$\mathcal{C}_{\IIind}(\boldsymbol{\sigma}, \mathbf{X})$. After such displacements, the
motion is unbounded because it must lead to a variation of the discrete
component of the relaxed spin~\cite{erementchoukSelfcontained2023}.

The second part of the theorem can be regarded as following from Earnshaw's
theorem about stability of a system of charges from
electrostatics~\cite{griffithsIntroduction2017} or can be proven directly
by respectively adapting the usual argument used to prove Earnshaw's
theorem. We start from a useful technical statement

\begin{lemma}\label{thm:weak-even-unstable-single}
  Let a weak generic cluster $\sset{V}^{(p)}$ occupying interval
  $\left( X_-^{(p)}, X_+^{(p)} \right)$, such that either $X_-^{(p)} > 0$
  or $X_+^{(p)} < 0$ (so that $\sset{V}^{(p)}$ does not contain the
  auxiliary spin) have sufficiently high charge,
  $\abs{Q^{(p)}(\boldsymbol{\sigma})} > 1$, where
  $Q^{(p)}(\boldsymbol{\sigma}) = \sum_{\alpha \in \sset{V}^{(p)}} q_\alpha$. Then
  $\sset{V}^{(p)}$ is unstable.
\end{lemma}
\begin{proof}
  According to Eq.~\eqref{eq:e_field-def-single}, the equations of motion
  governing the relaxed spins inside $\sset{V}^{(p)}$ are determined by the
  field $\epsilon(X_\alpha)$ that can be written as
  \begin{equation}\label{eq:weak_vp_field}
    \epsilon\left( X_\alpha \right)  = \epsilon^{(\bar{p})} + \epsilon^{(p)}\left( X_\alpha \right) ,
  \end{equation}
  where $\epsilon^{(\bar{p})}$ and $\epsilon^{(p)}\left( X_\alpha \right)$ are the
  contributions due to the relaxed spins inside and outside of
  $\sset{V}^{(p)}$, respectively. Since $\abs{Q^{(p)}} \geq 2$, there are two
  distinct relaxed spins in $\sset{V}^{(p)}$ with the charge $q^{(p)}$ of
  the same sign as $Q^{(p)}$ and extremal values of their $X$-coordinates
  among the relaxed spins in $\sset{V}^{(p)}$ with the same charge. Let
  these marginal relaxed spins be $\beta$ and $\gamma$, and
  $X_- < X_\beta < X_\gamma < X_+$. Then, we have
  \begin{equation}\label{eq:weak_vp_spreading}
    \frac{d}{dt} \left( X_\gamma - X_\beta \right)  = q^{(p)}
    \left[ \epsilon^{(p)}\left( X_\gamma \right) - \epsilon^{(p)}\left( X_\beta \right) \right]
    = 2 q^{(p)} \left[  q^{(p)} + \sum_{\alpha: X_\beta < X < X_\gamma} q_\alpha \right] > 0.
  \end{equation}
  The inequality follows from comparing the expression in the square
  brackets with the total charge of $\sset{V}^{(p)}$.
  
  Thus, the separation between $\beta$ and $\gamma$ increases leading to either
  unbound evolution or collision with other clusters. 
\end{proof}

\begin{proof} [Proof of Theorem~\ref{thm:solution-single}]
  The whole set of relaxed spins can be regarded as a weak cluster with a
  charge of even (including zero) magnitude. To apply
  Lemma~\ref{thm:weak-even-unstable-single}, we need to incorporate the
  auxiliary spin with its fixed continuous component.

  The argument used for proving Lemma~\ref{thm:weak-even-unstable-single}
  would be invalidated if the auxiliary spin is the only carrier of the
  excess charge. It is easy to show, however, that this is impossible.
  Without loss of generality, we can assume that the excess charge is
  positive, $Q(\boldsymbol{\sigma}) > 0$. Now, suppose the auxiliary spin with
  $q_0 > 0$ is the only positively charged particle, so that all free
  relaxed spins have $q_\alpha = -1$. However, in this case, the total charge is
  $q_0 + \sum_{\alpha \ne 0} q_\alpha = N - 2P - N = -2P \leq 0$, which contradicts the
  assumption that $Q(\boldsymbol{\sigma}) > 0$.

  What remains to be checked is how Eq.~\eqref{eq:weak_vp_spreading}
  changes when one of the marginal relaxed spins is the auxiliary spin. Let
  $Q(\boldsymbol{\sigma}) > 0$, and the auxiliary spin with $q_0 > 0$ and
  relaxed spin $\beta$ with $X_\beta > 0$, are positively charged relaxed spins
  with the smallest and the largest continuous component, respectively.
  Then, we have
  \begin{equation}\label{eq:v2_spread_aux}
    \frac{dX_\beta}{dt} =
    q_\beta \left[ \sum_{\alpha: X < X_\beta} q_\alpha - \sum_{\alpha: X > X_\beta} q_\alpha \right] \geq
    q_\beta \left[ Q(\boldsymbol{\sigma}) - q_\beta \right]  > 0,
  \end{equation}
  where we have taken into account that $\abs{Q(\boldsymbol{\sigma})}$ is even.
  Equation~\eqref{eq:v2_spread_aux} is intentionally written in the form
  valid for both $Q(\boldsymbol{\sigma}) > 0$ and $Q(\boldsymbol{\sigma}) < 0$ cases.
  It is elementary to adopt the argument to the case when $X_\beta < 0$ and to
  show that in this case $dX_\beta / dt < 0$.

  Thus, generic weak clusters can only converge to a strong cluster if
  they represent a solution of the tomography problem.
\end{proof}

We conclude the analysis of the single-ray case by noting that except for
the cases with all pixels set or unset, single-ray tomography problems have
multiple solutions~\cite{erementchoukSelfcontained2023}. In this case, the
terminal state can be expected to consist of multiple strong clusters.
Since Lemma~\ref{thm:weak-even-unstable-single} is applicable to each
cluster independently, we obtain that the resultant strong clusters must
have zero charge. In particular, this implies that if a strong cluster does
not contain the auxiliary spin, then, generically, it should comprise only
two relaxed spins.


\subsection{Terminal states for multiple rays}

A generalization of the approach in the previous section to tomography
problems with $R > 1$ rays requires accounting for the fact that the
relaxed spins belonging to different rays do not interact with each other.
This is achieved by associating with the relaxed spins vector charges
defined in the $R$-dimensional Euclidean space. To this end, in
$\mathbb{R}^R$, we introduce basis vectors $\vec{e}(r)$ with coordinates
$\vec{e}(r)_j = \delta_{r,j}$. Next, for each relaxed spin
$\alpha$ we introduce
$\sset{D}(\alpha) = \left\{ r : \alpha \in \sset{R}(r) \right\}$, the set of rays
containing $\alpha$. Finally, we define the vector charge for a free relaxed
spin $\alpha \ne 0$
\begin{equation}\label{eq:vect_charge_free_def}
  \vec{q}_\alpha = \sigma_\alpha \sum_{r \in \sset{D}(\alpha)} \vec{e}(r).
\end{equation}
Respectively, for the auxiliary spin, we define
\begin{equation}\label{eq:vect_charge_aux_def}
  \vec{q}_0 = \sum_{r \in \sset{D}(\alpha)} \widetilde{q}_0(r)\vec{e}(r).
\end{equation}

Using these definitions, one can define vector charges of collections of
relaxed spins as a sum of vector charges. For example, the total vector
charge carried by state $\left( \boldsymbol{\sigma}, \mathbf{X} \right)$ is
$\vec{Q}(\boldsymbol{\sigma}) = \sum_{\alpha \in \widetilde{\sset{V}}} \vec{q}_\alpha$, while
the total charge of relaxed spins in ray $r$ is natural to define as
$\vec{Q}(r; \boldsymbol{\sigma}) = \sum_{\alpha \in \sset{R}(r)} \vec{q}_\alpha$. In
terms of vector charges, the cut weight of graph
$\graph{G}_{\mathbf{s}, \mathbf{P}}$ [Eq.~\eqref{eq:cut_tom_graph}] takes a
simple form
\begin{equation}\label{eq:cut_tom_graph_vect_charges}
  \mathcal{C}_{\graph{G}_{\mathbf{s}, \mathbf{P}}}(\boldsymbol{\sigma}) = 
  \overline{\mathcal{C}}_{\graph{G}_{\mathbf{s}, \mathbf{P}}} -
  \frac{1}{4} \vec{Q}(\boldsymbol{\sigma}) \cdot \vec{Q}(\boldsymbol{\sigma}),
\end{equation}
where $\vec{a} \cdot \vec{b}$ denotes the inner product in Euclidean space.
Equation~\eqref{eq:cut_tom_graph_vect_charges} is obtained directly from
Eq.~\eqref{eq:cut_tom_graph} by noticing that $Q(r; \boldsymbol{\sigma}) =
\vec{e}(r) \cdot \vec{Q}(\boldsymbol{\sigma}) =
\vec{e}(r) \cdot \vec{Q}(r; \boldsymbol{\sigma})$.  

Respectively, the relaxed cut function [Eq.~\eqref{eq:cut-v2} written for
the discrete tomography problem] is written as
\begin{equation}\label{eq:cut-v2-DT}
  \mathcal{C}_{\IIind}(\boldsymbol{\sigma}, \mathbf{X}) =
  \mathcal{C}_{\graph{G}_{\mathbf{s}, \mathbf{P}}}(\boldsymbol{\sigma}) +
  \frac{1}{4} \sum_{\alpha, \beta} \vec{q}_\alpha \cdot \vec{q}_\beta \abs{X_\alpha - X_\beta}.  
\end{equation}

Finally, the equations of motion governing the \mdII{} model are a
straightforward generalization of Eqs.~\eqref{eq:eqm-v2-q-single}
and~\eqref{eq:e_field-def-single}: 
\begin{equation}\label{eq:eqm-v2-q-multi}
  \dot{X}_\alpha = \vec{q}_\alpha \cdot \vec{\epsilon}\left( X_\alpha \right) ,
\end{equation}
where
\begin{equation}\label{eq:e_field-def-multi}
  \vec{\epsilon}\left( X \right) = \sum_{\beta:X_\beta < X} \vec{q}_\beta - \sum_{\beta:X_\beta > X} \vec{q}_\beta.
\end{equation}

A substantial technical deviation from the single-ray case is expressed by
the fact that in the vector case, the formal equilibrium corresponds not
only to the case of vanishing vector field but also to the case of
orthogonal $\vec{q}_\alpha$ and $\vec{\epsilon}\left( X_\alpha \right)$. Dynamically, the
orthogonality expresses the exact cancellation of the contributions
originating from different rays containing the relaxed spin. Such
cancellation invalidates an important property self-evident in the
single-ray case that in equilibrium there are no isolated relaxed spins.

The accidental orthogonality can be eliminated by ensuring that projections
$\vec{e}(r) \cdot \vec{Q}(r)$ for different $r$ are linearly independent over
$\left\{ -1, 1 \right\}$. In our simulations, such linear independence was
forced by redefining the last term in Eq.~\eqref{eq:cut-v2-DT} as
$\frac{1}{4} \vec{Q}(\boldsymbol{\sigma}) \cdot \Lambda \cdot \vec{Q}(\boldsymbol{\sigma})$, where
$\Lambda = \mathrm{diag}(\lambda(1), \ldots, \lambda(R))$ is a diagonal matrix with positive
linearly independent $\lambda(r) > 0$. Alternatively, this procedure can be
understood as redefining $\vec{e}(r) \to \sqrt{\lambda(r)}\vec{e}(r)$. In
computations, it was sufficient to generate the diagonal elements as
$\lambda_r = \sqrt{p(r)} / \lfloor\sqrt{p(r)}\rfloor$, where
$\left\{ p(r) \right\}$ is a set of random distinct prime numbers.

Satisfying the tomographic data for the $r$-th ray results in
$\vec{e}(r) \cdot \vec{Q}(r) = 0$. This provides a way for translating the
results obtained in the previous section to the multi-ray case and
identifying the main barriers for converging to a
reconstructed image starting from generic initial conditions and the
dynamical features of the \mdII{} model that allow overcoming these
barriers.

\begin{figure}[tb]
  \centering
  \includegraphics[width=0.75\textwidth]{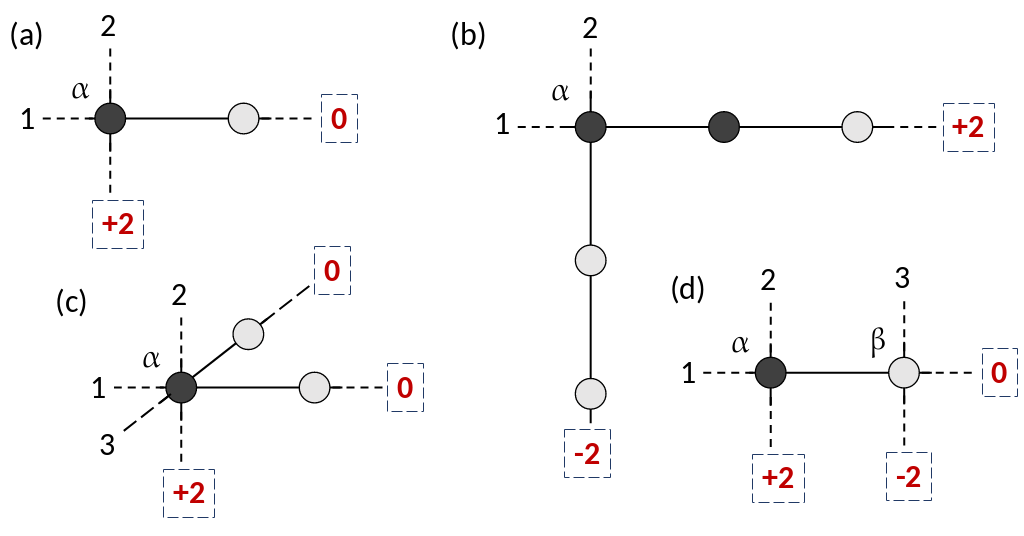}
  \caption{Examples of configurations violating the tomographic data that
    are stable equilibria for the relaxed spin $\alpha$. Dashed-framed numbers
    show deviations of the respective components of the vector charge from
    zero due to mismatched tomographic data. Circles depict states of the
    discrete components that need to be corrected. (a) Spin $\alpha$ is in
    neutral equilibrium. (b) Spin $\alpha$ is asymptotically stable (there is a
    returning field acting on $\alpha$). Five spins need to be corrected to
    match the tomographic data. (c) A configuration similar to~(a) but at
    the intersection of three rays. Spin $\alpha$ is asymptotically stable
    state, while only three spins need to be corrected. (d) A realization
    of the configuration from~(a) for a tomography problem on a 2D grid
    leads to violating the tomographic data in two rays. Both $\alpha$ and
    $\beta$ are in neutral equilibrium.}
  \label{fig:alpha-stable}
\end{figure}

The conclusion that the \mdII{} states
$\left( \boldsymbol{\sigma}, \mathbf{X} \right)$ with $\boldsymbol{\sigma}$
satisfying the tomographic data are the only stable terminal states remains
valid for the problems with multiple rays as well. As discussed above, this
follows from the fact that the relevant critical points of
$\mathcal{C}_{\IIind}(\boldsymbol{\sigma}, \mathbf{X})$ besides max-cut are saddle points.

On the other hand, the argument used for proving the second part of
Theorem~\ref{thm:solution-single} is invalid. This reflects the fact that
for multiple-ray tomography problems, Earnshaw's-like statements about
instability of \emph{individual} charges are no longer true for vector
charges. As demonstrated by Fig.~\ref{fig:alpha-stable}, there do exist
arrangements of vector charges stable with respect to any single-particle
disturbances. Respectively, to match the tomographic data in the presence
of these patters requires inversion of \emph{multiple} spins. Consequently,
local search-based approaches to solve the spin formulation of the discrete
tomography problem are prone to terminate in states with such defects as
manifested in numerical simulations shown in
Section~\ref{sec:random-images}.

In turn, inherently, the \mdII{} model's trajectories are not constrained
to Hamming-limited variations of the discrete component of the relaxed
spin~\cite{erementchoukSelfcontained2023}. As a result, the \mdII-machine
can either avoid defects or escape them after the random agitation of a
terminal state breaking the tomographic data.

To discuss this property of the \mdII{} model, we limit ourselves to the
tomography problem on a 2D rectangular grid, where the typical defects in
terminal states stable with respect to single-particle perturbations have
the form shown in Fig.~\ref{fig:alpha-stable}. It is not difficult to see
that such patterns lead to mismatching the tomographic data in, at least,
two rays.

\begin{lemma}\label{thm:two-rays-lemma}
  For a tomographic problem on a rectangular grid, if the spin
  configuration $\boldsymbol{\sigma}$ does not match the tomographic data,
  $\vec{Q}(\boldsymbol{\sigma}) \cdot \vec{Q}(\boldsymbol{\sigma}) > 0$, then there are
  at least two rays with unsatisfied tomographic data.
\end{lemma}

\begin{figure}[tb]
  \centering
  \includegraphics[width=0.75\textwidth]{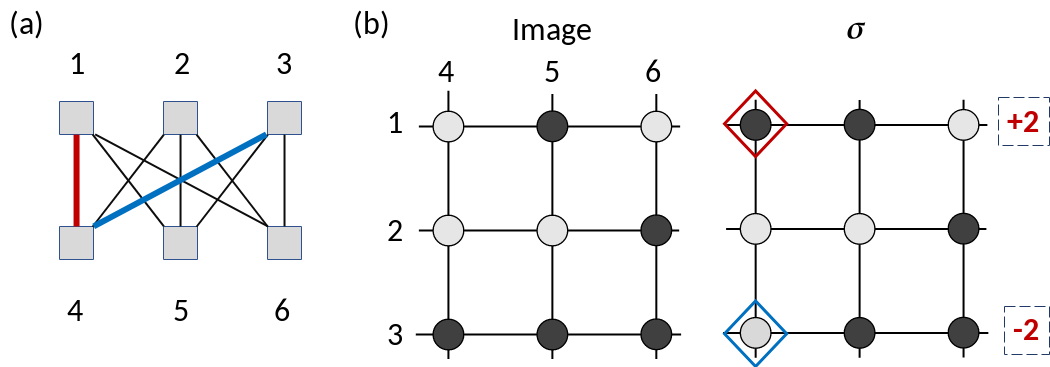}
  \caption{(a) An example of the ray graph $\sset{G}_{\sset{R}}$ for the
    discrete tomography problem on the $3\times3$ grid. The highlighted edges
    show edge-induced subgraph
    $\sset{T}_{\sset{R}}(\boldsymbol{\sigma}' \vert \boldsymbol{\sigma}) \subseteq
    \sset{G}_{{\sset{R}}}$. (b) An example of a $3\times3$ image and an incorrect
    spin configuration stable with respect to single-spin perturbations.
    Correction of the selected spins is represented by
    $\sset{T}_{\sset{R}}(\boldsymbol{\sigma} \vert \boldsymbol{\sigma})$ shown in~(a).}
  \label{fig:ray-graph}
\end{figure}

\begin{proof}
  We represent the system of
  rays by graph $\sset{G}_{\sset{R}}$, whose nodes correspond to rays, and
  two nodes are connected if the respective rays intersect. For example,
  the graph representing the tomography problem for a $W \times W'$ image is a
  complete bipartite graph $\sset{K}_{W, W'}$ with partitions,
  $\sset{A}$ and $\sset{B}$, corresponding to rows and columns,
  respectively (see
  Fig.~\ref{fig:ray-graph}). We notice that the edges in this graph
  correspond to the image pixels. Indeed, two rays intersect only at a
  single pixel, and each pixel lies at the intersection of at most two
  rays, thus there is a bijection between pixels and the edge set of
  $\sset{G}_{\sset{R}}$. Having such correspondence established, we assign
  to the edges weights $\pm 1$ according to the value of the respective
  spins. In other words, we regard the spin configuration
  $\boldsymbol{\sigma}$ as defining a function
  $\boldsymbol{\sigma} : \sset{E}\left( \sset{G}_{\sset{R}} \right) \to \left\{
    -1, 1 \right\}$.

  Let the tomographic constraint be broken in ray $r_1$ by having one too
  many set pixels, so that
  $Q(r_{1}; \boldsymbol{\sigma}) = \vec{e}(r_1) \cdot \vec{Q}(r_1; \boldsymbol{\sigma}) =
  2$. Let the error be correctable by inverting relaxed spin
  $\alpha \in \sset{R}(r_1)$ with $\sigma_\alpha = 1$. This relaxed spin lies at the
  intersection of rays $\sset{R}(r_1)$ and $\sset{R}(r_2)$ and, by
  assumption, $Q(r_2; \boldsymbol{\sigma}) = 0$. Thus, there another
  relaxed spin $\beta \in \sset{R}(r_2) \cap \sset{R}(r_{3})$ with
  $\sigma_\beta = -1$ must be inverted together with $\sigma_\alpha$ to maintain zero error in
  $\sset{R}(r_2)$. Consequently, there exists relaxed spin
  $\gamma \in \sset{R}(r_{3}) \cap \sset{R}(r_{4})$ with
  $\sigma_{\gamma} = 1$ that must be inverted to maintain zero error in
  $\sset{R}(r_{3})$, and so on.

  This observation shows that a transformation from the configuration
  $\boldsymbol{\sigma}$ to a correct configurations $\boldsymbol{\sigma}'$ selects a
  subset of edges in $\sset{G}_{\sset{R}}$ corresponding to the set of
  spins that need to be inverted. This subset defines an edge-induced
  subgraph
  $\sset{T}_{\sset{R}}\left(\boldsymbol{\sigma} \right) \subseteq \sset{G}_{\sset{R}}$.
  For each node
  $\sset{R}(r) \in \sset{T}_{\sset{R}}(\boldsymbol{\sigma}' \vert \boldsymbol{\sigma})$,
  the total variation of $Q(r; \boldsymbol{\sigma})$ is equal
  to the doubled total weight of the edges incident to $\sset{R}(r)$.
  Since, as was discussed above, the tomographic data may not uniquely
  define the image, there may be multiple graphs
  $\sset{T}_{\sset{R}}\left(\boldsymbol{\sigma} \right)$ describing the
  respective correcting transformations. Since, by the main assumption, the
  tomography problem has a solution, there exists at least one such graph,
  which is sufficient for our purposes.

  First, we notice that if the degree of $\sset{R}(r)$ for some $r$ is odd,
  then $\abs{Q(r; \boldsymbol{s})} > 0$. But there must be an even number
  of nodes with odd degree in a graph, implying the lemma.

  It remains to consider the case when
  $\sset{T}_{\sset{R}}(\boldsymbol{\sigma})$ is Eulerian. It is worth
  noticing that due to the even degrees, they correspond to correcting
  configurations $\boldsymbol{\sigma}$ with strong
  violations of the tomographic constraint, $\abs{Q(r; \boldsymbol{\sigma})} \geq
  4$, which, as follows from the proof of
  Lemma~\ref{thm:weak-even-unstable-single}, cannot emerge from weak clusters.

  For Eulerian $\sset{T}_{\sset{R}}(\boldsymbol{\sigma})$ the multiplicity of
  rays mismatching the tomographic data follows from the fact that that
  there are no odd cycles in $\sset{T}_{\sset{R}}(\boldsymbol{\sigma})$. To
  employ this property, we introduce an orientation of edges in
  $\sset{T}_{\sset{R}}(\boldsymbol{\sigma})$. Without loss of generality, we
  assume that the tomographic constraint is broken in ray
  $\sset{R}(r_{1}) \in \sset{A}$ and $Q(r_{1}; \boldsymbol{\sigma}) > 0$. Then, we
  define edges with positive and negative weights as oriented from
  $\sset{A}$ to $\sset{B}$ and from $\sset{B}$ to $\sset{A}$, respectively.
  Then, due to the absence of odd cycles, any directed cycle defines such a
  set of relaxed spins whose inversion does not change the ray charges and,
  therefore, cannot correct the mismatched tomographic data. Such directed
  cycles can be removed without loosing the property of
  $\sset{T}_{\sset{R}}(\boldsymbol{\sigma})$ to describe a correcting
  transformation.

  Next, we construct a directed trail starting from $\sset{R}(r)$ following
  the orientation of the incident edges. Any return to $\sset{R}(r)$
  defines a directed cycle, which can be removed from
  $\sset{T}_{\sset{R}}(\boldsymbol{\sigma})$ without exhausting the edges
  outgoing from $\sset{R}(r)$. Without directed cycles, a directed trail
  starting from $\sset{R}(r)$ must terminate at another ray $\sset{R}(r')$
  with an imbalance between incoming and outgoing edges and, hence,
  $\abs{Q(r'\boldsymbol{\sigma})} > 0$.

  It is worth noticing that the case of Eulerian
  $\sset{T}_{\sset{R}}(\boldsymbol{\sigma})$ corresponds to strong violations of
  the tomographic constraint, $\abs{Q(r\boldsymbol{\sigma})} \geq 4$ and
  $\abs{Q(r'\boldsymbol{\sigma})} \geq 0$. Adapting the argument used in the proof
  of Lemma~\ref{thm:weak-even-unstable-single}, it can be shown that the
  respective configurations $\boldsymbol{\sigma}$ cannot emerge from a weak
  cluster.
\end{proof}

In view of Lemma~\ref{thm:two-rays-lemma}, we limit ourselves to
considering the simplest case of a defect shown in
Fig.~\ref{fig:alpha-stable}(d) and showing that random agitations can lead
to rearranging of both spins $\alpha$ and $\beta$ eliminating errors in both rays.
It is not difficult to see that it is sufficient that after the random
agitation, relaxed spins $\alpha$ and $\beta$ both have, say, positive continuous
components maximal compared to other relaxed spins in rays containing $\alpha$
and $\beta$. Indeed, in this case, the equations of motion of these relaxed
spins take the form
\begin{equation}\label{eq:ab-escape}
  \begin{split}
    \frac{dX_\alpha}{dt} & =
                      \vec{q}_\alpha \cdot
                      \left( \vec{Q}(\boldsymbol{\sigma}) - \vec{q}_\alpha -
                      \vec{q}_\beta \right) +
    \vec{q}_\alpha \cdot \vec{q}_\beta \sgn \left( X_\alpha - X_\beta \right), \\ 
    \frac{dX_\beta}{dt} & =
                      \vec{q}_\beta \cdot
                      \left( \vec{Q}(\boldsymbol{\sigma}) - \vec{q}_\alpha -
                      \vec{q}_\beta \right) +
    \vec{q}_\alpha \cdot \vec{q}_\beta \sgn \left( X_\beta - X_\alpha \right).
  \end{split}
\end{equation}
Taking into account that $\vec{Q}(\boldsymbol{\sigma}) = 2\left( \vec{q}_\alpha +
  \vec{q}_\beta \right)$, we find
\begin{equation}\label{eq:ab-escape-2}
  \frac{d}{dt}\left( X_\alpha + X_\beta \right) = \left( \vec{q}_\alpha + \vec{q}_\beta
  \right)^2 > 0.
\end{equation}
Considering other particles in the rays containing $\alpha$ and $\beta$ with the
$X$-coordinates taking the maximal (after excluding $\alpha$ and $\beta$) value, one
can easily check that they are non-increasing. Thus, $\alpha$ and $\beta$ remain
outsiders with increasing $X$-coordinates, hence their motion is unbound.
For the \mdII{} model evolution, this implies that both spins $\alpha$ and $\beta$
will invert, which eliminates the error in recovering the image.

\section{Numerical demonstrations}
\label{sec:numerics}

The theoretical framework developed above shows the origin of the the
\mdII{} model capability to solve discrete tomography problem. At the same
time, solving practical tomography problems requires much more than
recovering \emph{an} image from tomographic data. For example, one may
impose additional constraints on the recovered image (see,
e.g.~\cite{gardnerDiscreteTomographyDetermination1997}) or accounting for
possible errors in the tomographic data
itself~\cite{cekoErrorCorrectionDiscrete2022}, and so forth. 
In the present paper, we limit ourselves to demonstrations of the basic
capabilities of the \mdII{} model on two simplified examples:
reconstructing 2D and 3D images. The evolution of the \mdII{} model was
simulated using the Euler approximation supplied with the rule of updating
the binary component of the relaxed spin as illustrated in
Fig.~\ref{fig:spin_phase_space}. While the Euler approximation is not
trouble-free and may lead to emergence of various
artifacts~\cite{shuklaNonbinary2024}, it is sufficient for our purposes in
the present paper. In all the examples, the \mdII-machine ran unsupervised
without any additional processing and only random agitations were applied.

\subsection{Tomography of 2D binary images}
\label{sec:random-images}

As the first demonstration, we studied the probability $P_{\mathrm{succ}}$
of successful reconstruction of an image from its tomographic data on a set
of random square binary images with the sides varying from $4$ to $30$. For
each size, five images were generated with the probability $p = 0.5$ of a
pixel to be activated. For both approaches, $P_{\mathrm{succ}}$ was
obtained by averaging over the set of five images.

\begin{figure}[tb]
  \centering
  \includegraphics[width=0.95\textwidth]{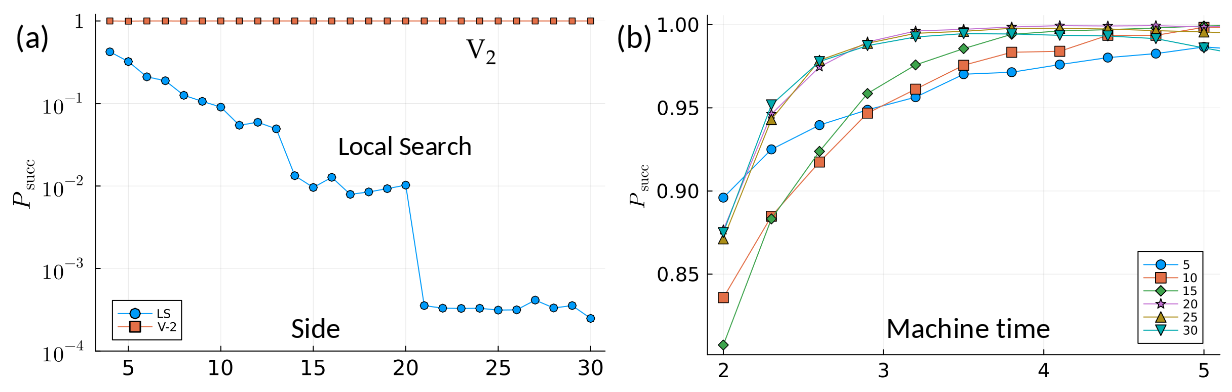}
  \caption{Probability of successful reconstruction of images
    $P_{\mathrm{succ}}$ from the tomographic data for square random binary
    images. (a) Comparison of $P_{\mathrm{succ}}(W)$ as a function of the
    image side length, $W$, for two approaches: using the \mdII{} model and
    $1$-opt local search. (b) The variation of $P_{\mathrm{succ}}$ with the
    duration of the free evolution stage between random agitations for
    random images of selected sizes: $5, 10, 15, 20, 25, 30$.}
  \label{fig:p_v_N}
\end{figure}

In results shown in Fig.~\ref{fig:p_v_N}, the \mdII-machine time measured
in the inverse units used in the equations of
motion~\eqref{eq:eqm-v2-q-multi} and was limited to $T = 5.0$ between
agitations for the results shown in Fig.~\ref{fig:p_v_N}(a) (the number of
agitations was limited to $10$) and changed in the interval $[2.0, 5.3]$
for Fig.~\ref{fig:p_v_N}(b) (the number of agitations was fixed to $5$).
The wall-time required by a hardware of a software implementation to
process one agitation depends on such details as the duration of the time
steps in the Euler approximation and organization of computation. The
simulations used a fixed number, $M = 600$, of time steps with the duration
of $\Delta t = \frac{T}{M}$. For each time-step, the computations used the
sparsity of the graph $\graph{G}_{\mathbf{s}, \mathbf{P}}$. Thus, the time
complexity of the \mdII{} simulations shown in Fig.~\ref{fig:p_v_N} can
be estimated as $\mathcal{O}(MW^3)$, where $W$ is the width of the field, per one
agitation stage. An efficient parallelization of computations can be
expected to reduce the wall-time scaling to $\mathcal{O}(MW)$ per agitation stage.

Figure~\ref{fig:p_v_N}(a) compares two approaches to recovering an image
from its tomographic data using the spin-based formulation of the
tomography problem: using the \mdII{} model and $1$-opt local search
(greedy search) that ensures that in the terminal state the cut weight
cannot be improved by inverting any single spin. The success probability
was sampled for the \mdII{} model by restarting it from $100$ random
initial conditions. Since the successful recovery by the local search turns
with increasing the image size into a rare event, for the local search,
$P_{\mathrm{succ}}$ was sampled by restarting it from $10^6$ random initial
states. While studying the performance of local search on a spin
representation of the discrete tomography problem is not our objective, we
would like to notice a curious behavior of $P_{\mathrm{succ}}(W)$: it drops
significantly when the most probable total mismatch with the tomographic
data increases by two, or the number of defects like shown in
Fig.~\ref{fig:alpha-stable}(d) increases by one, but changes only
moderately between these transitions.

Figure~\ref{fig:p_v_N}(a) demonstrates the inherent complexity of the
discrete tomography problem. Its spin representation does not determine
whether an image can be successfully recovered from its
tomographic data by a particular implementation of the Ising machine. The
ability of the \mdII{} model to solve the tomography problem is a result of
its specific dynamical properties, as discussed in the previous sections.

To provide a better illustration of how the dynamics of the \mdII{} model
affects the solution of the tomography problem, Fig.~\ref{fig:p_v_N}(b)
shows the dependence of $P_{\mathrm{succ}}$, obtained from sampling
terminal states obtained from $10^3$ random initial conditions, on duration
of the evolution between random agitations. This dependence demonstrates
the emergence of several universal dynamical features, such as the
universal convergence of $P_{\mathrm{succ}}$ to $1$ at $T \approx 3.5$, whose
investigation is a subject of ongoing research. It should be noted that as
the number of time steps covering the whole interval between the random
agitations was kept constant, $M = 600$, for all situations, the errors
related to the duration the size of the time step become apparent for
larger images at $T \gtrsim 5.0$.

\subsection{Tomography of 3D binary images}
\label{sec:3d-pcb}


As a demonstration of advanced capabilities of the \mdII{} dynamics, we
consider reconstruction of a multi-layered printed circuit board. This
problem has a straightforward connection with the general discrete
tomography problem, as any PCB (printed circuit board) without integrated
circuit chips and other discrete elements can be modelled a
three-dimensional binary matrix composed of elements
$\left\{ 0, 1 \right\}$, where ‘1’ represents the presence of conducting
material. We restrict the scale of this demonstration so that the number of
required spins is in the order of $10^3$ and hence, consider the
reconstruction of small enough $4$-layered PCB, such that the maximum size
of the projection image is $16 \times 16$ pixels. Moreover, we assume that the
ray projection data is available along only three dimensions of the object.

Figure~\ref{fig:3d-PCB}a shows the section of an artificial 4-layered PCB
we consider for reverse engineering with the conducting tracks depicted in
yellow and the non-conducting base in brown. Figure~\ref{fig:3d-PCB}b shows
the mock tomographic data of the PCB along in the $XY$, $YZ$, and $ZX$
planes. We obtain this by summing up the discrete binary matrix modelling
the PCB in Fig.~\ref{fig:3d-PCB}a along three dimensions.
Figure~\ref{fig:3d-PCB}c shows reconstructed PCB layers.

\begin{figure}[tb]
  \centering
  \includegraphics[width=0.95\textwidth]{3d-PCB.png}
  \caption{(a) External view of a 4-layered PCB; (b) model tomographic data
    collected along $X$, $Y$ and $Z$ dimensions; (c) reconstructed PCB
    layers }
  \label{fig:3d-PCB}
\end{figure}

\section{Conclusion}

We investigated the capability of a relaxation-based dynamical Ising
machine driven by the \mdII{} dynamical model to solve the discrete
tomography problem about recovering a binary image from its tomographic
data, the pixel sum along a discrete set of rays.

We show that in the simplest single-ray case, the convergence of the
\mdII{} model is governed by the same principles as stability of a system
of electrostatically interacting charged particles in 1D. Based on this
observation, we show that the \mdII{} model terminates in a solution to the
tomography problem starting from generic initial conditions.

We generalized this approach to the multiple-ray case by identifying a
vector charge with the relaxed spins. We identified the main bottleneck
terminal states that do not solve the tomography problem, but the \mdII{}
may converge to. We show that random agitations can move the system away
from these states, and, as the numerical simulations show, only a few such
agitations are needed to ensure the convergence with a high probability
($P_{\mathrm{succ}} \approx 1$). Two key properties of the \mdII{} model are
responsible for this. First, random agitations do not worsen the spin
configuration in the terminal state of the system of the relaxed spins.
Second, spin transitions occurring during the evolution of the \mdII{}
model are not Hoffman-limited. This should be contrasted to local
search-based treatments of the spin reformulation of the discrete
tomography problem, for which such transitions present a strong barrier and
lead to the quick deterioration of performance with the problem size
manifested in numerical simulations.

Our consideration demonstrates that specific dynamical models can be
approached as unconventional algorithms producing exact solutions to
highly nontrivial data processing tasks. This advances the conventional
approach to dynamical Ising machines, regarding them as heuristic
devices targeting approximate solutions to optimization problems.

\section*{Acknowledgements}
The work was partially supported by the US National Science Foundation
(NSF) under Grant No. 2531175 and partially No. 1909937.

\bibliography{tomo.bib}

\end{document}